\documentclass[aps,prl,twocolumn,showpacs,superscriptaddress,groupedaddress]{revtex4-1}
\usepackage{graphicx}
\usepackage{amssymb}
\usepackage{amsmath}
\usepackage{latexsym}
\usepackage{ulem}
\usepackage{color}
\usepackage{dcolumn}
\usepackage{subfigure}
\usepackage[T1]{fontenc}
\usepackage[utf8]{inputenc}
\usepackage[english,frenchb]{babel}
\usepackage[colorlinks=true,linkcolor=blue,citecolor=blue]{hyperref}%

\usepackage{bm}        
\usepackage{amssymb}   
\begin{document}

\title{Absence of confinement in (SrTiO$_{3}$)/(SrTi$_{0.8}$Nb$_{0.2}$O$_{3}$) superlattices}

\author{G. Bouzerar}
\email[E-mail:]{georges.bouzerar@univ-lyon1.fr}
\affiliation{Univ Lyon, Universit\'e Claude Bernard Lyon 1, CNRS, Institut Lumi\`ere Mati\`ere, F-69622, LYON, France}
\author{S. Th\'ebaud}
\affiliation{Univ Lyon, Universit\'e Claude Bernard Lyon 1, CNRS, Institut Lumi\`ere Mati\`ere, F-69622, LYON, France}
\author{R. Bouzerar}
\affiliation{Univ Lyon, Universit\'e Claude Bernard Lyon 1, CNRS, Institut Lumi\`ere Mati\`ere, F-69622, LYON, France}
\author{S. Pailh\`es}
\affiliation{Univ Lyon, Universit\'e Claude Bernard Lyon 1, CNRS, Institut Lumi\`ere Mati\`ere, F-69622, LYON, France} 
\author{Ch. Adessi}
\affiliation{Univ Lyon, Universit\'e Claude Bernard Lyon 1, CNRS, Institut Lumi\`ere Mati\`ere, F-69622, LYON, France}                          
\date{\today}
\selectlanguage{english}
\begin{abstract}
The reduction of dimensionality is an efficient pathway to boost the performances of thermoelectric materials, it leads to the quantum confinement of the carriers and thus
to large Seebeck coefficients (S) and it also suppresses the thermal conductivity by increasing the phonon scattering processes. However, quantum confinement in superlattices is not always easy to achieve and needs to be carefully validated. In the past decade, large values of S have been measured in (SrTiO$_{3}$)/(SrTi$_{0.8}$Nb$_{0.2}$O$_{3}$) superlattices (Nat. Mater. $\textbf{6}$, 129 (2007) and Appl. Phys. Lett. $\textbf{91}$, 192105 (2007)). In the $\delta$-doped compound, the measured S was almost 6 times larger than that of the bulk material. This huge increase has been attributed to the two dimensional confinement of the carriers in the doped regions. In this work, we demonstrate that the experimental data can be well explained quantitatively within the scenario in which electrons are delocalized in both in-plane and growth directions, hence strongly suggesting that the confinement picture in these superlattices may be unlikely. 
\end{abstract}
\maketitle

Over the past two decades, and because of increasing energy and environmental issues, thermoelectric materials have re-gained a great interest owing to their ability 
to convert waste heat into electricity \cite{Snyder,Disalvo,Dresselhaus,Nolas,book,Heremans,Tritt}. Typically the performance of a thermoelectric device is controlled by the dimensionless thermoelectric figure of merit $ZT=\frac{S^{2}\sigma T}{\kappa}$ where $S$ is the Seebeck coefficient, $\sigma$ is the electrical conductivity, T is the temperature and $\kappa=\kappa_e + \kappa_{ph}$ is the total thermal conductivity which contains both electronic and phonon contributions. Thus a good thermoelectric material requires a large Seebeck, a high electrical conductivity, and a low thermal conductivity.
In the recent years, most efforts to improve ZT have focused on reducing the lattice thermal conductivity by enhancing the phonon scattering processes. To achieve the reduction of $\kappa$, there are different efficient modus operandi such as alloying \cite{Steele,Cahill}, increasing the anharmonicity \cite{Garg}, or even by introducing nanoinclusions/inhomogeneties into the bulk matrix \cite{Faleev,Wang,Zhang}. Since the first thermoelectric Bi$_2$Te$_3$ alloy has been discovered, the room-temperature figure of merit of bulk semiconductors has increased only marginally.
However, recent studies in nanostructured thermoelectric materials have opened interesting pathways towards materials exhibiting large ZT \cite{Dresselhausb, Mahan, Mingo, Farhangfar, Zhou}.
The main ideas behind the nanostructuration are twofold. First, it leads to the quantum confinement of the carriers, inducing sharp peaks in the density of states, and therefore giving rise to a simultaneous increase of both the Seebeck coefficient and the electrical conductivity. Second, the nanostructuration suppresses the thermal conductivity by increasing the phonon scattering. 
This strategy has for instance been applied to thin films superlattices such as Bi$_2$Te$_3$/Sb$_2$Te$_3$ \cite{Venka1,Venka2}, quantum dot superlattices PbSeTe/PbTe \cite{Harman} and bulk alloys BiSbTe \cite{Poudel}. However, achieving quantum confinement of the itinerant carriers in superlattices is not a simple and straightforward task \cite{Vineis}. As an example, it has been claimed in Ref. \onlinecite{Harman2} that the strong enhancement (with respect to the bulk compound) of both the Seebeck coefficient and the Figure of merit in PbSeTe/PbTe quantum dot superlattices originated from the quantum confinement of the carriers. It has been shown later \cite{Harman3}, that the carrier densities (Hall measurements) were actually incorrect leading to a wrong interpretation of the measured Seebeck coefficients. Therefore it has been concluded that this PbSeTe/PbTe superlattice did not exhibit any confinement. Thus, experimental measurements that do not constitute a direct probe of the confinement effects should be analysed carefully.

Recently, it has been argued that resulting from the two dimensional confinement of the electrons in (SrTiO$_{3}$)$_x$/(SrTi$_{0.8}$Nb$_{0.2}$O$_{3}$)$_y$ superlattices ($x$ and $y$ are respectively the number of undoped and Nb doped layers), giant Seebeck coefficients have been measured \cite{Otha1,Otha2}. In particular, in the extreme limit of a single Nb doped layer ($y=1$ and $x$ varies), the measured value of S as a function of $x$ could saturate at values almost 6 times larger than that of the bulk material. In addition, it has been concluded that the critical barrier thickness for quantum electron confinement was about 6.25 nm (16 unit cells of STO). However, it is important to notice that the large increase of the Seebeck coefficient does not provide a direct signature of the 2D quantum confinement. In this work, we demonstrate that the measured data could be explained (qualitatively and quantitatively) assuming the absence (or weakness) of quantum confinement in these superlattices. 

\begin{figure*}[t]\centerline
{\includegraphics[width=7 in,angle=0]{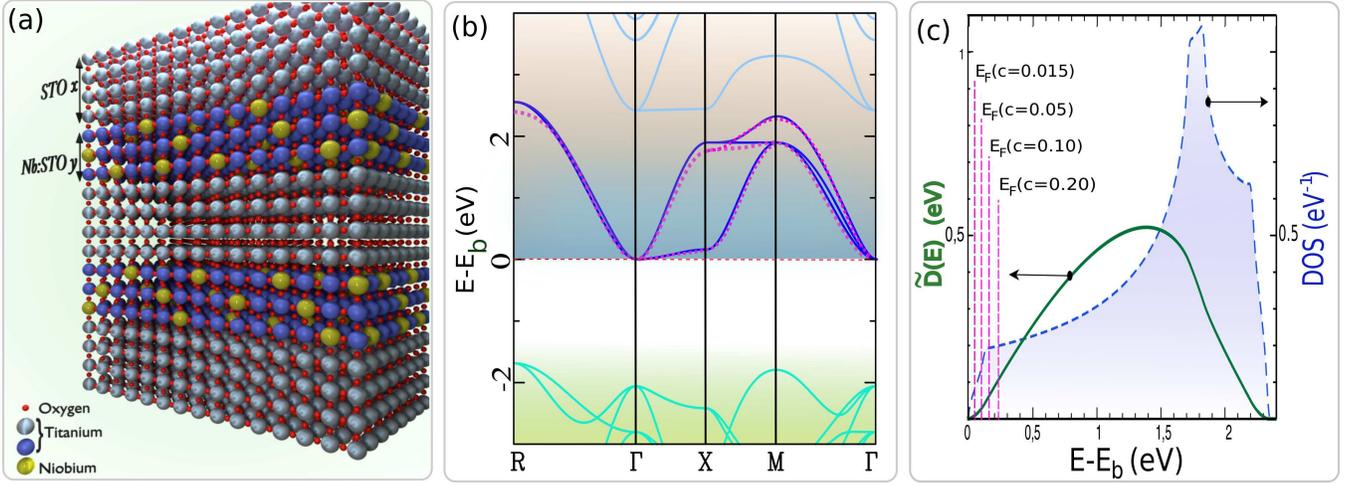}}
\caption{(Color online)
(a) Schematic view of the super-lattice structure (STO)$_x$(STO:Nb)$_y$ (b) Bulk band structure from {\textit ab initio} SIESTA (continuous lines) and from the minimal tight binding (TB) model (pink dashed lines).The green lines correspond to the valence band and the blue lines to the conduction bands. (c) TB calculations of the bulk density of states and reduced Drude weight $\overset{\sim}{D}(E)=\frac{\hbar}{\sigma_0} D(E)$, E$_b$ is the energy of the bottom of the conduction band. The vertical dashed lines indicate the position of the Fermi level for various concentration of carriers.
}
\label{fig1}
\end{figure*} 

In a recent study, we have shown that the thermoelectric properties of electron doped STO (conductivity and Seebeck coefficient) could be well understood and reproduced within the framework of a realistic Tight Binding (TB) Hamiltonian (3 t$_{2g}$ bands) that includes electron-electron scattering mechanism and disorder treated in the Born approximation \cite{Bouzerar}. The hopping integrals of the TB Hamiltonian were directly extracted from ab initio based studies. The Hamiltonian reads, $\hat{H_{0}}=\sum_{\textbf{k},\alpha} \epsilon^{0}_{\alpha}(\textbf{k}) c_{\footnotesize\textbf{k}\alpha}^{\dagger}c_{\footnotesize\textbf{k}\alpha}$ where $\alpha$ denotes the orbital index. The d$_{xy}$ band dispersion is $\epsilon^{0}_{xy}(\textbf{k})=-2t_{1}\left(cos(k_{x}a)+cos(k_{y}a)\right)-2t_{2}cos(k_{z}a)-4t_{3}cos(k_{x}a)cos(k_{y}a)$, where $a$ is the lattice parameter. The two other bands (d$_{yz}$ and d$_{zx}$) are obtained straightforwardly by a circular permutation of $(x,y,z)$. The hopping parameters obtained from Wannier projections are t$_{1}$=0.277 eV, t$_{2}$=0.031 eV and t$_{3}$=0.076 eV as estimated in Ref. \onlinecite{3dtbmodel}. The formalism is further detailed in Ref. \onlinecite{Bouzerar}.

We now propose to briefly summarize the procedure that allows to calculate the Seebeck coefficient as a function of $x$, $y$ (number of undoped and doped layers respectively), and temperature T. The calculations will be directly compared to the existing and available experimental data. The conductivity and the Seebeck coefficient are given by, 
\begin{eqnarray}
\sigma(\mu,T)=-\int \Sigma(E,T)\frac{\partial f}{\partial E}dE,
\end{eqnarray}
\begin{eqnarray}
S(\mu,T)=\frac{1}{eT\sigma(\mu,T)}\int \Sigma(E,T)(E-\mu)\frac{\partial f}{\partial E}dE,
\end{eqnarray}
where $\mu$ is the T-dependent chemical potential, $f$ is the Fermi-Dirac distribution, $\Sigma(E,T)=D(E)\tau(E,T)$ is the Transport Distribution Function where D(E) is the Drude weight (calculated at T=0 K) and $\tau(E,T)$ is the energy and temperature dependent quasiparticle lifetime.
We restrict ourselves to the weak disorder regime, a well justified approximation for samples exhibiting a good metallic behaviour. This regime corresponds to $k_{F}l_{e} \gg$ 1, where $k_{F}$ is the Fermi wave vector and $l_{e}$ the mean free path. In this regime, D(E) can be well approximated by $D(E) \approx -\frac{\sigma_{0}}{N\hbar}\langle \hat{K_x} \rangle(E)$, where $\sigma_{0}=\frac{e^2}{\hbar a}= 6258~\Omega^{-1}\cdot cm^{-1}$, N is the total number of sites and $\hat{K_x}=-\frac{\partial^{2}\hat{H_{0}}}{\partial \kappa_{x}^{2}}$ ($\kappa_{x}=k_{x}a$).
It is worth mentioning that in the $x$-direction, $D(E)$ is dominated by the d$_{xy}$ and d$_{xz}$ bands that contribute equally, whilst the d$_{yz}$ band has a negligible contribution. The hopping integral in the $x$-direction is indeed very small in the later case. In Fig.~\ref{fig1} we have plotted, (a) a schematic view of the super-lattice structure, (b) the bulk band structure from ab initio (SIESTA)\cite{Siesta} and from the minimal TB model and (c) the TB calculations for the bulk density of states and reduced Drude weight. The position of the Fermi level for various carrier concentrations per unit cell is also shown.

\begin{figure}[t]\centerline
{\includegraphics[width=1.0\columnwidth,angle=0]{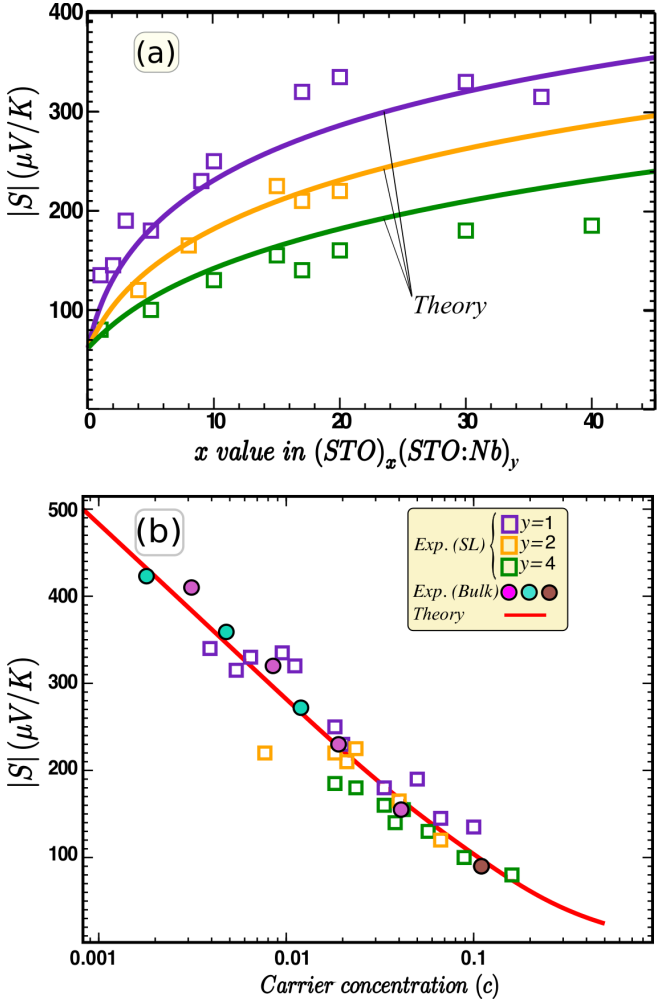}}
\caption{(Color online) (a) Seebeck coefficient at T=300 K in the super-lattice (STO)$_x$(STO:Nb)$_y$, values of $y$ are indicated in bottom panel and $x$ varies from 0 to 50. Open squares are the experimental data from Ref. \onlinecite{Otha2}, the continuous lines are the TB calculations. (b) S as a function of $c=\frac{y}{x+y} c_D$, where $c_D=0.20$ (Nb concentration in the doped regions). The bulk experimental data are extracted from Refs. \onlinecite{Stemmer,Otha-bulk,Jalan}.
The continuous line corresponds to the TB calculations.
}
\label{fig2}
\end{figure} 
The Transport Distribution Function requires both the energy dependent Drude weight and the electron lifetime $\tau(E,T)$. $\tau(E,T)$ has two contributions: $\frac{1}{\tau(E,T)} = \frac{1}{\tau_{dis}(E)}+\frac{1}{\tau_{th}(T,E)}$. $\tau_{dis}(E)$ denotes the effect of disorder resulting from the cationic substitutions and presence of other defects (intrinsic, dislocations, grain boundaries) whilst $\tau_{th}(T,E)$ is the temperature dependent part. Its origins are electron-electron (e-e) scattering and electron-phonon (e-ph) scattering. In oxides such as STO, several studies showing a $T^2$ dependent resistivity suggest that the e-e mechanism prevails over the e-ph contribution up to relatively large temperatures \cite{Baratoff,VanderMarel1,Mikheev,Klimin}. This has been confirmed by the good agreement found between theory and experiment in Ref. \onlinecite{Bouzerar}. From the Fermi golden rule we assume $\frac{\hbar}{\tau_{dis}(E)}= \frac{\pi W^2}{6} \rho(E)$ where $\rho(E)$ is the density of states and $W$ measures the strength of the disorder. The $T$ and $E$ dependent contribution is assumed to have the form, $\frac{\hbar}{\tau_{th}(E)}=C\frac{(k_{B}T)^{2}}{E-E_{b}}$ where $C$ is a dimensionless constant and $E_{b}$ the energy at the bottom of the conduction band. For electron doped STO it was shown that ($C$=24.5, $W$=0.17 eV) allows to describe the physics quantitatively for a wide range of doping and dopants for both the Seebeck coefficient and the electrical conductivity. Note also that the strength of the disorder is small compared to the bandwidth which is of the order of 2.5 eV (see Fig.~\ref{fig1}), thus it validates the weak disorder approximation.

We now consider the scenario in which there is no 2D quantum confinement of the electrons in the superlattices (STO)$_{x}$(STO:Nb)$_y$. In order to calculate the Seebeck coefficient we assume a uniform carrier density per unit cell in the overall superlattice. The Nb concentration per unit cell in the doped regions in the measured samples is $c_{D}=0.20$ (Ref. \onlinecite{Otha2}). To allow the direct comparison with the experimental data, our calculations are performed assuming a uniform carrier density per unit cell $c=\frac{y}{x+y} c_D$. It is important to stress that from now on, our theory is completely free of fitting parameters. We would like to emphasize that for the temperature range considered here (300 K to 900 K), the Seebeck coefficient is almost independent from both $C $ and $W$. Thus, the only relevant physical ingredients are (i) the details and accuracy of the band structure and (ii) the form of $\tau_{th}(T,E)$. If the electrons are really confined in the growth direction in these superlattices, we should expect a strong disagreement between our calculations and the experimental measurements, that would completely invalidate our procedure.

In Fig.~\ref{fig2}(a) the Seebeck coefficient at T=300 K in the super-lattice (STO)$_x$(STO:Nb)$_y$ is shown as a function of $x$ for $y=$ 1, 2 and 4. We clearly observe an overall good agreement between the measured values and the calculated ones. In Fig.~\ref{fig2}(b) the data are now plotted as a function of $c=\frac{y}{x+y} c_D$, we find that the experimental data are well reproduced by the theoretical curve that assumes a uniform distribution of the carriers in the superlattice. The experimental data points exhibit some dispersion that may reflect the quality of the samples, the presence of native defects such as oxygen vacancies, dislocations, interface defects/deformations, sample history and also the fact that the Nb concentration may fluctuate from sample to sample.

\begin{figure}[t]\centerline
{\includegraphics[width=1.0\columnwidth,angle=0]{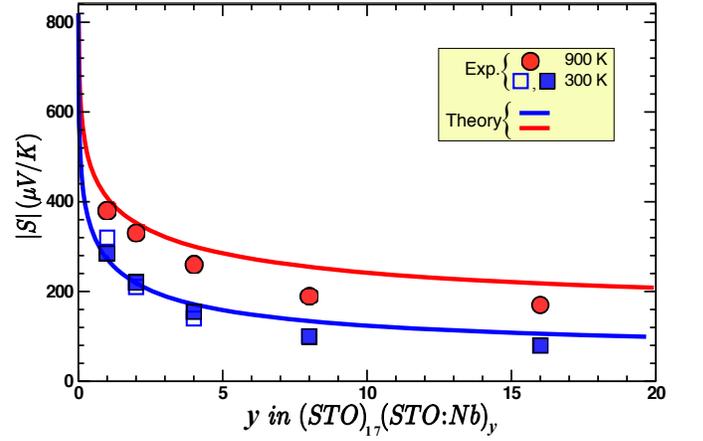}}
\caption{(Color online) 
Seebeck coefficient at $T=$300 K and 900 K in the super-lattice (STO)$_{x=17}$(STO:Nb)$_y$ as a function of $y$. Symbols are the experimental data from Ref. \onlinecite{Otha2,Otha-pssb} 
and the continuous lines are the TB calculations.
}
\label{fig3}
\end{figure}

\begin{figure}[t]\centerline
{\includegraphics[width=1.0\columnwidth,angle=0]{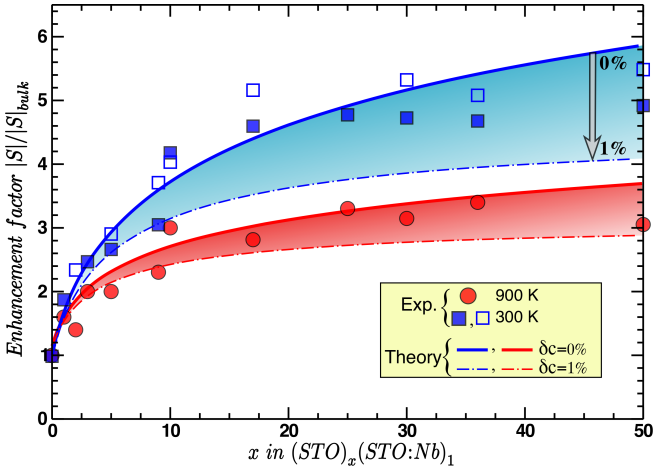}}
\caption{(Color online) Enhancement factor $\lvert \frac{S}{S_{Bulk}}\rvert$ of the Seebeck coefficient in the super-lattice (STO)$_{x}$(STO:Nb)$_1$ as a function of $x$ for both $T=$300 K and 900 K.
The filled region (blue for $T=$300 K  and red for $T=$ 900 K) indicate the effects of an additional concentration of carriers $\delta c $ up to 1$\%$. The experimental data (symbols) are extracted from Refs \onlinecite{Otha2,Otha-pssb} 
}
\label{fig4}
\end{figure}

In Fig.~\ref{fig3} we now focus on the effect of the thickness of the Nb doped region assuming a fixed value for the undoped one. We have now plotted the Seebeck coefficient as a function of $y$ for two different temperatures, namely T=300 K  and 900 K, the number of undoped layers is set to $x=17$. As we increase $y$ the amplitude of the Seebeck coefficient decreases as a consequence of the increase of the overall carrier density. We again find a good agreement between the theory of a uniformly distributed electron gas and the experimental data, this agreement is even excellent at room temperature.
In addition, we expect a saturation of the Seebeck coefficient for large $y$ at S$_{Bulk}$(c$_D$), which appears to be the case already for $y=20$.

In the next figure, Fig.~\ref{fig4}, we now focus on the particular case of a $\delta$-doped compound, $y=1$ for which a large increase of the Seebeck has been reported in Ref. \onlinecite{Otha2,Otha-pssb}. We plot the enhancement factor $\lvert \frac{S}{S_{Bulk}}\rvert$ as a function of $x$ and for two different temperatures (T= 300 K and 900 K) where $S_{Bulk}$ refers to the 20\% doped bulk material that corresponds to $x=0$. 
As mentioned above, the density of carriers can not be precisely tuned experimentally, as a result of various mechanisms. Indeed, as seen from Hall measurements in Ref. \onlinecite{Otha2}, the density of electrons per doped layer can fluctuate by as much as 30$\%$ from sample to sample. Therefore, we include the effects of these variations by adding typically 1\% additional carriers per unit cell.
To be more specific, we perform the calculations for $c=\frac{y}{x+y} c_D + \delta c$ with $\delta c$ up to 1\% per unit cell. Note that performing realistic calculations including defects such as oxygen vacancies or dislocations would be extremely complicated and demanding (requires extremely large supercells) and would go beyond the scope of the present manuscript. Let us now discuss the results. First notice that the experimental data, for a given value of $x$, are sample sensitive especially for large $x$ (see full squares and empty squares), the enhancement factor can vary by about 20$\%$. More generally, there is some dispersion in the experimental data, especially strong around $x=10$. However, the agreement between theory and experiments is rather good, and even better at large temperature. As expected, the effect of additional carriers becomes more pronounced as we increase $x$. Thus, if $\delta c$ is constant, it would result in the saturation of the enhancement factor at large $x$ but it should be noticed, that no critical or characteristic length-scale can be extracted.

\begin{figure}[t]\centerline
{\includegraphics[width=1.0\columnwidth,angle=0]{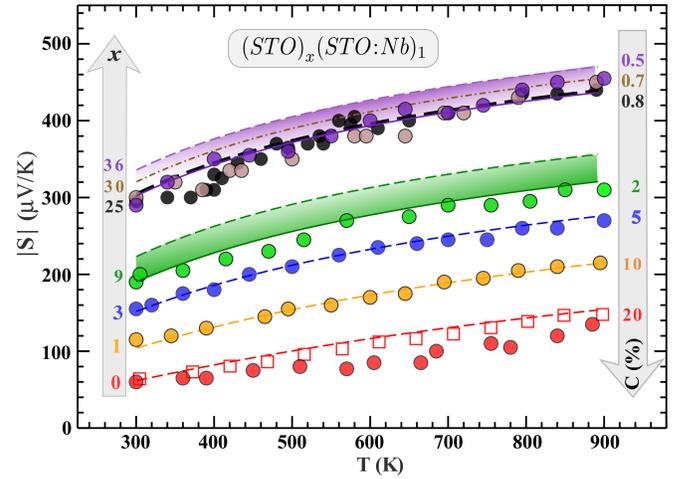}}
\caption{(Color online) Seebeck coefficient as a function of temperature (from T= 300 K to 900 K) in the super-lattice (STO)$_{x}$(STO:Nb)$_1$ ($y$ is fixed), where $x$ varies from 0 (20$\%$ doped bulk material) to 36. Filled and open Symbols are experimental data from Ref. \onlinecite{Otha-pssb} and Ref. \onlinecite{Sonne} respectively. The dashed and continuous lines are the TB calculations.
The shaded regions correspond to the effect of $\delta c$ up to 1\% for $x=$ 9 (green region) and $\delta c$ up to 0.25\% for $x=$ 36 (purple region). The effective concentration $c=\frac{y}{x+y} c_D$ is also plotted in the figure.
}
\label{fig5}
\end{figure} 

We now study the effect of temperature (it varies from T= 300 K to 900 K) on the Seebeck coefficient S in the super-lattice (STO)$_{x}$(STO:Nb)$_1$, where $x$ ranges from 0 (20$\%$ doped bulk material) to 36. The results are depicted in Fig.~\ref{fig5}. First, regarding the bulk data ($x=0$) we observe that the theory agrees very well with the data from Ref. \onlinecite{Sonne}. The measured bulk data of Ref. \onlinecite{Otha-pssb} are slightly smaller and appear to fluctuate with the temperature. Note that, for these data, the calculations would fit better assuming a slightly larger Nb concentration of the order of 23\% instead of 20\%. On the other hand, our calculated Seebeck coefficient agrees perfectly well with the experimental data for both $x=1$ and $x=3$, for the overall range of temperature. For $x=9$ the calculated Seebeck coefficients are slightly larger (by 10-15\%). However, assuming a small additional amount of electrons ($\delta c= 1\%$ only), the agreement between theory and experiment now becomes excellent. Note also, that adding a small concentration of electrons for both $x=1$ and $x=3$ would only weakly affect the results (the average concentration would only weakly change). Regarding larger values of $x$ ($x=25$, 30 and 36) we first notice that the experimental data strongly fluctuates with the temperature, the average carrier concentrations in these superlattices are relatively low: 0.8\%, 0.7\% and 0.5\% respectively. For $\delta c= 0$ the agreement is good but the theoretical Seebeck coefficients are still slightly larger.However, by adding just 0.25\% of carriers, the agreement between theory and experiment becomes excellent.

To conclude, we have demonstrated that the recently reported giant increase of the Seebeck coefficients in (SrTiO$_{3}$)$_x$/(SrTi$_{0.8}$Nb$_{0.2}$O$_{3}$)$_y$ superlattices is fully consistent with the absence of 2D quantum confinement of the carriers in the doped regions. Our conclusion is further supported by the observation that the power factor ($\sigma S^2$) measured in these superlattices is close to that of the bulk electron doped STO \cite{Otha3}.
It would be of great interest to confirm whether our scenario is correct by direct measurements such as transverse resistivity, Angle Resolved Photoemission Spectroscopy, or by a direct probe of the depth profile of the carrier concentration. Oxide based thermoelectric superlattices are promising materials for high ZT devices but achieving a true 2D quantum confinement requires (i) a suitable choice of dopant that has a drastic effect on the host band structure in the vicinity of the Fermi level or (ii) a more appropriate choice for the undoped compound.

\begin{acknowledgments}

\end{acknowledgments}

\end{document}